\documentstyle[preprint,aps]{revtex}

\begin{document}
\preprint{quant-ph/9504009}

\title{Discrete photodetection and Susskind-Glogower ladder
operators}
\author{Y. Ben-Aryeh and C. Brif
\footnote{e-mail: phr65bc@phys1.technion.ac.il}}
\address{Department of Physics, Technion -- Israel Institute of
Technology, Haifa 32000, Israel}
%\date{}
\maketitle

\begin{abstract}

We assert that state reduction processes in different types of
photodetection experiments are described by using different kinds
of ladder operators. A special model of discrete photodetection
is developed by the use of superoperators which are based on the
Susskind-Glogower raising and lowering operators.

\end{abstract}
\newpage
%\vspace*{1cm}

In order to give a quantum description of optical phase, Susskind
and Glogower (SG) \cite{SG} have defined in the harmonic oscillator
Hilbert space the following raising and lowering operators:
	\begin{equation}
\hat{E}_{+} = \sum_{n=0}^{\infty} |n+1 \rangle\langle n| ,
\mbox{\hspace{1cm}}
\hat{E}_{-} = \sum_{n=0}^{\infty} |n \rangle\langle n+1| ,
\mbox{\hspace{1cm}}
\hat{E}_{-} |0\rangle = 0 .
\label{1}
	\end{equation}
Here $|n\rangle$ $(n=0,1,\ldots ,\infty)$ are the number states.
However, since the number spectrum is restricted from below by the
vacuum state, the SG operators $\hat{E}_{+}$ and $\hat{E}_{-}$ are
not unitary \cite{SG,CN}:
	\begin{equation}
\hat{E}_{-} \hat{E}_{+} = \hat{1} ,  \mbox{\hspace{1cm}}
\hat{E}_{+} \hat{E}_{-} = \hat{1} - |0 \rangle\langle 0| .
\label{2}
	\end{equation}
This nonunitarity leads to difficulties in the quantum description
of optical phase \cite{SG,CN}. This problem can be solved either by
using a finite-dimensional Hilbert space \cite{PB} or by using the
antinormal ordering of the SG operators $\hat{E}_{+}$ and
$\hat{E}_{-}$ in the usual infinite-dimensional Hilbert space
\cite{LuPer,BBA}. In the present work we would like to discuss
another aspect concerning the SG operators $\hat{E}_{+}$ and
$\hat{E}_{-}$. We will show that these operators can describe a state
reduction process in a sort of discrete photodetection of the
single-mode radiation fields represented by the density operators
diagonal in the number state representation.

Usually, photodetection of the single-mode radiation field is
described by the use of the mode annihilation
and creation operators $\hat{a}$ and $\hat{a}^{\dagger}$, which
can be written in terms of the SG operators and the number operator
$\hat{n} = \sum_{n=0}^{\infty} n |n \rangle\langle n|$ in the
following form
	\begin{equation}
\hat{a} = \sqrt{ \hat{n} +1 } \hat{E}_{-} , \mbox{\hspace{1cm}}
\hat{a}^{\dagger} = \hat{E}_{+} \sqrt{ \hat{n} +1 } .
\label{3}
	\end{equation}
When $\hat{a}$ and $\hat{a}^{\dagger}$ lower or raise a number state
$|n\rangle$, they also generate the weight factor $\sqrt{n}$ or
$\sqrt{n+1}$, respectively. The SG operators $\hat{E}_{+}$ and
$\hat{E}_{-}$ only raise or lower the number states
without generating any weight factor. This essential difference
between the two types of ladder operators implies differences
between photodetection schemes, in whose descriptions different
types of ladder operators are used.

A usual model of continuous photodetection is the so-called
closed-system model [6-9], in which both
the radiation field and the photodetector are enclosed in a cavity,
and the measurement is continuous. The density operator of the field
is continuously reduced by the information provided by the
photodetector. The instantaneous process of one-photon counting
is described by the superoperator ${\cal J}$:
	\begin{equation}
\hat{\rho}(t^{+}) = {\cal J} \hat{\rho}(t) \equiv
\frac{ \hat{a} \hat{\rho}(t) \hat{a}^{\dagger}
}{ {\rm Tr}\, [ \hat{\rho}(t) \hat{a}^{\dagger} \hat{a} ] } .
\label{4}
	\end{equation}
Here $\hat{\rho}(t)$ and $\hat{\rho}(t^{+})$ are the density
operators for the radiation field immediately before and after the
detection. The superoperator ${\cal J}$ consists of nonunitary
transformation (describing state reduction) and the normalization.
The no-count process which occurs for a duration time $\tau$ is
described by the superoperator ${\cal S}_{\tau}$:
	\begin{equation}
\hat{\rho}(t+\tau) = {\cal S}_{\tau} \hat{\rho}(t) \equiv
\frac{ \exp(-\frac{1}{2}\lambda\hat{a}^{\dagger}\hat{a}\tau)
\hat{\rho}(t) \exp(-\frac{1}{2}\lambda\hat{a}^{\dagger}\hat{a}\tau)
}{ {\rm Tr}\, [ \hat{\rho}(t)
\exp(-\frac{1}{2}\lambda\hat{a}^{\dagger}\hat{a}\tau) ] } .
\label{5}
	\end{equation}
Here $\lambda$ is a parameter characteristic of the coupling between
the detector and the field. For a measurement, where an $n$-photon
state is converted to an $(n-1)$-photon state whenever a photon is
detected, the photodetection probability is proportional to $n$.
This proportionality is described by the use of $\hat{a}$ and
$\hat{a}^{\dagger}$ in equations (\ref{4}) and (\ref{5}) \cite{Lee}.

We suggest another photodetection scheme in which the single-mode
radiation field is enclosed in a cavity. We send two-level Rydberg
atoms in the lower state through the cavity, one after another, and
measure their states at the exit. This experimental scheme is
similar, from the technical point of view, to a micromaser \cite{MM}.
However, we propose to use this system for realizing a special kind
of photodetection. We use only radiation fields whose density
operators are diagonal in the number state representation,
	\begin{equation}
\hat{\rho} = \sum_{n=0}^{\infty} p(n) |n \rangle\langle n| .
\label{6}
	\end{equation}
When the measurement shows that one atom is excited, it means that
one photon is subtracted from the radiation field. The detection of
an excited atom is the only referred process in this model. Our idea
is that in this photodetection scheme the field reduction is
described by the superoperator ${\cal B}_{-}$ which includes the SG
operators:
	\begin{equation}
\hat{\rho}_{-1} = {\cal B}_{-} \hat{\rho} \equiv
\frac{ \hat{E}_{-} \hat{\rho} \hat{E}_{+} }{
1 - \langle 0|\hat{\rho}|0 \rangle }
\label{7}
	\end{equation}
where $\hat{\rho}$ and $\hat{\rho}_{-1}$ are the density operators
for the radiation field before and after the subtraction of a photon.
The normalization factor is ${\rm Tr}\, ( \hat{\rho} \hat{E}_{+}
\hat{E}_{-} ) = 1 - \langle 0|\hat{\rho}|0 \rangle$. In order to
understand why equation (\ref{7}) is valid we must show the
differences between our model of discrete photodetection and the
closed-system model of continuous photodetection. In discrete
photodetection the measurement occurs only when an atom leaves the
cavity, so that the number of measurements is equal to the number
of atoms transmitted through the cavity. At that, the only referred
measurement is that in which an excited atom is detected. Therefore
in our model there is no analog to the no-count process of continuous
photodetection. In continuous photodetection the measurement occurs
at any time whenever the photodetector is active in the cavity, and
the one-photon counting is referred as well as the no-count process.
There the measurement is made inside the cavity by the interaction of
the field with the detector. In our model the interaction between the
field and the atoms is inside the cavity but the detector measures
the states of the atoms outside the cavity, i.e., the detection is
separated from the interaction with the field. Although the
interaction in the cavity depends on the number of photons, the
information obtained by us (the excitation of an atom) is independent
of the features of interactions inside the cavity. The idea is that
we are not interested in the properties of interactions inside the
cavity and in the associated probabilities, it is of no concern to us
how many atoms in the lower state we must send to obtain one of them
in the excited state at the exit. By getting only the information
that one atom is excited we reduce an $n$-photon state of the
radiation into an $(n-1)$-photon state in a way that is independent
of $n$. This independence is described by the use of $\hat{E}_{+}$
and $\hat{E}_{-}$ in equation (\ref{7}).

When the field is in the number state $|n\rangle$, the field
reductions according to equations (\ref{4}) and (\ref{7}) are
equivalent. Our model cannot be applied to the field states given by
quantum mixtures of number states, $\sum_{n=0}^{\infty} C_{n}
|n\rangle$. By getting only the information that one photon is
absorbed we cannot conclude how the amplitudes $C_{n}$ are changed.
However, our model can be used for statistical mixtures of the form
(\ref{6}). For a state described by the density operator of the form
(\ref{6}), we have statistical probability $p(n)$ that the state is
$|n\rangle$ but in fact only one of the states $|n\rangle$ exists in
the cavity. As the result of state reduction (\ref{7}), the changes
in the photon-number distribution of the radiation field can be
expressed in our model in the following form
	\begin{equation}
p_{-1}(n) = \langle n| \hat{\rho}_{-1} |n \rangle =
\frac{ \langle n| \hat{E}_{-} \hat{\rho} \hat{E}_{+} |n \rangle }{
1 - \langle 0|\hat{\rho}|0 \rangle } = \frac{ p(n+1) }{ 1 - p(0) } .
\label{8}
	\end{equation}
For comparison, the continuous photodetection model gives
for the one-count process
	\begin{equation}
p(n,t^{+}) = \langle n| \hat{\rho}(t^{+}) |n \rangle =
\frac{ \langle n| \hat{a} \hat{\rho}(t) \hat{a}^{\dagger} |n \rangle
}{ \langle \hat{n} \rangle_{t} } = \frac{ n+1 }{ \langle \hat{n}
\rangle_{t} } p(n+1,t) .
\label{9}
	\end{equation}

The use of Bayes theorem \cite{Fri,Lee} enables to obtain these
results in a way that clarifies the principal differences between
the two models. Bayes theorem can be written in the form
	\begin{equation}
P(B_{j}|A) = \frac{P(B_{j})P(A|B_{j})}{\sum_{j} P(B_{j})P(A|B_{j})} ,
\label{10}
	\end{equation}
where $P(B|A)$ is the conditional probability that event $B$ occurs
under the condition that event $A$ is known to have occurred, and the
mutually exclusive events $B_{j}$ span the whole sample space:
$\sum_{j} P(B_{j}) = 1$. Let event $A$ be the detection of a photon
and $B_{j}$ be the fact that there is a certain number of photons
in the cavity. In the continuous photodetection the probability that
one of $n$ photons in the cavity is detected during the time $dt$
is $n \lambda d t$ and that no photon is detected is $(1 - n \lambda
d t)$. Therefore Bayes theorem (\ref{10}) can be written, after
it is known that one photon is detected at time $t$, as \cite{Lee}
	\begin{equation}
p(n,t^{+}) = \lim_{dt \rightarrow 0} \frac{ p(n+1,t) (n+1) \lambda
d t }{ \sum_{n=1}^{\infty} p(n,t) n \lambda d t } =
\frac{ n+1 }{ \langle \hat{n} \rangle_{t} } p(n+1,t) .
\label{11}
	\end{equation}
This equation is identical to equation (\ref{9}). In our model the
photodetection process occurs only when an atom is measured to be
in the excited state, i.e., we refer only to the information that one
photon is subtracted from the cavity. In this type of experiment,
where we wait any time till we observe an excited atom, the
probability $P(A)$ is equal to $1$. Then Bayes theorem (\ref{10}) can
be written, after it is known that one photon was subtracted from the
cavity, as
	\begin{equation}
p_{-1}(n) = \frac{ p(n+1) }{ \sum_{n=1}^{\infty} p(n) } =
\frac{ p(n+1) }{ 1 - p(0) } .
\label{12}
	\end{equation}
This equation is identical to equation (\ref{8}).

The mean photon number immediately after the measurement of an
excited atom can be easily calculated in our model by using
equation (\ref{8}) or (\ref{12}) for the photon-number distribution.
We get
	\begin{equation}
\langle \hat{n} \rangle_{-1} = \sum_{n=0}^{\infty} n p_{-1}(n) =
\frac{ \langle \hat{n} \rangle }{ 1 - p(0) } - 1 .
\label{13}
	\end{equation}
The denominator $1 - p(0)$ takes into account the fact that it is
impossible to excite a photon from the vacuum. If the field was
initially in the vacuum state, there is no photodetection process
in our model and the number of photons in the cavity remains zero.
With the exception of the vacuum-dependent factor, the mean photon
number is merely reduced by 1 after the detection of an excited atom.
The situation in the closed-system model of continuous photodetection
is quite different. By using equation (\ref{9}) or (\ref{11}), one
obtains the mean photon number immediately after the one-count
process:
	\begin{equation}
\langle \hat{n} \rangle_{t^{+}} = \sum_{n=0}^{\infty} n p(n,t^{+}) =
\langle \hat{n} \rangle_{t} - 1 + \frac{ (\Delta n)^{2}_{t} }{
\langle \hat{n} \rangle_{t} } .
\label{14}
	\end{equation}
Here the photon-number variance is defined by $(\Delta n)^{2} =
\langle \hat{n}^{2} \rangle - \langle \hat{n} \rangle^{2}$. This
result shows that the the mean photon number of the post-measurement
state depends on the pre-measurement photon statistics \cite{Ueda}.
The difference between the mean photon numbers before and after the
one-count process is not exactly equal to 1, but it has an additional
term depending on the photon-number variance before the measurement.

We can generalize our model by sending atoms in the lower state
through the cavity till the measurement shows a desired number $N$ of
excited atoms. It means that $N$ photons were subtracted from the
cavity. Then the field state is reduced according to
	\begin{equation}
\hat{\rho}_{-N} = {\cal B}_{-}^{N} \hat{\rho} \equiv
\frac{ \hat{E}_{-}^{N} \hat{\rho} \hat{E}_{+}^{N} }{
 {\rm Tr}\, ( \hat{\rho} \hat{E}_{+}^{N} \hat{E}_{-}^{N} ) } .
\label{15}
	\end{equation}
Our experimental scheme also enables us to add photons to the cavity
(this process is inverse to photodetection). In this case we send
atoms in the upper state through the cavity, one after another, and
measure their states at the exit till the measurement shows a desired
number $N$ of de-excited atoms. It means that $N$ photons were added
to the cavity. Then the field state is reduced according to
	\begin{equation}
\hat{\rho}_{+N} = {\cal B}_{+}^{N} \hat{\rho} \equiv
\hat{E}_{+}^{N} \hat{\rho} \hat{E}_{-}^{N} .
\label{16}
	\end{equation}
It is interesting to note that the transformation (\ref{16}) is
unitary and there is no need for a normalization factor. We find
here a very special case where state reduction is described by a
unitary transformation. The transformation (\ref{15}) will be also
unitary for the density operator $\hat{\rho}$ obeying the following
condition
	\begin{equation}
p(n) = \langle n| \hat{\rho} |n \rangle = 0
\mbox{\hspace{0.4cm} for \hspace{0.4cm}} n < N .
\label{17}
	\end{equation}
The mechanism described by equation (\ref{16}) and by equation
(\ref{15}) under the condition (\ref{17}) is number shifting,
and we can refer to ${\cal B}_{+}^{N}$ and ${\cal B}_{-}^{N}$ as
number-shifter superoperators. Then we have an analogy between the
number shifter described by transformations (\ref{16}) and (\ref{15})
and the well known phase shifter described by the unitary
transformation
	\begin{equation}
\hat{\rho}_{\phi} = e^{i\phi\hat{n}} \hat{\rho} e^{-i\phi\hat{n}}
\label{18}
	\end{equation}
where phase shift $\phi$ is a real parameter.

In any real experiment we cannot ignore losses inside the cavity,
and the detector of the atoms is never perfect. These experimental
limitations introduce statistical features and thus destroy the
state reduction mechanism which is based on the exact information.
For imperfect detection we can generalize our model by assuming that
the measurement reduces the density operator into the form
	\begin{equation}
\hat{\rho}_{\pm\bar{N}} = \sum_{N} \alpha_{N}
{\cal B}_{\pm}^{N} \hat{\rho} .
\label{19}
	\end{equation}
The detector efficiency distribution $\alpha_{N}$ must be
sufficiently narrow around the true number $\bar{N}$ of excited (or
de-excited) atoms in order to realize our model.

In conclusion, in the present work we have developed a special kind
of discrete photodetection which is applicable to the single-mode
radiation fields represented by the density operators diagonal in the
number state representation. In this photodetection model state
reduction is described by superoperators which are based on the SG
raising and lowering operators.

%\pagebreak

\begin{flushleft}

\end{flushleft}

\end{document}